\documentclass[authoryear,preprint,review,12pt]{elsarticle}

\usepackage{amssymb}
\usepackage{amsmath,epsfig,wasysym,textcomp}
\usepackage{graphicx,graphics}
\usepackage{url}
\usepackage{hyperref}
\usepackage[capitalise]{cleveref}
\crefname{subsection}{subsection}{subsections}
\usepackage{textcomp}
\usepackage{enumitem}
\usepackage{float,rotating,pdflscape}
\usepackage{verbatim}
\usepackage{amsthm,amstext,mathtools}

\journal{Journal of Atmospheric and Solar-Terrestrial Physics}

\begin{document}

\begin{frontmatter}



\title{Impact of 15 Jan 2010 annular solar eclipse on the equatorial and low latitude ionosphere over Indian region from Magnetometer, Ionosonde and GPS observations}


\author[iit]{S. K. ~Panda\corref{cor1}\fnref{fn1}}
\ead{sampadpanda@gmail.com}

\author[iit]{S. S. ~Gedam}
\ead{shirish@iitb.ac.in}

\author[iigold,label4]{G. ~Rajaram}
\ead{girijarajaram@gmail.com}

\author[iignew]{S. Sripathi}
\ead{sripathi@iigs.iigm.res.in}

\author[iignew]{A. Bhaskar}
\ead{ankushbhaskar@gmail.com}
\cortext[cor1]{Corresponding author}
\fntext[fn1]{Telephone: +91 022 25764658, Fax: +91 022 25783190}
\address[iit]{Centre of Studies in Resources Engineering,
Indian Institute of Technology Bombay, Mumbai, India, Pin-400076.}
\address[iigold]{Prof.(Retd), Indian Institute of Geomagnetism, Mumbai, India, Pin-400005.}
\address[iignew]{Indian Institute of Geomagnetism, Navi Mumbai, India, Pin-410218.}
\address[label4]{Presently associated with Centre of Studies in Resources Engineering,
Indian Institute of Technology Bombay, Mumbai, India, Pin-400076.}

\begin{abstract}
The annular eclipse of Jan 15, 2010 over southern India was studied with a network of multi-instrumental observations consisting magnetometer, ionosonde and GPS receivers. By selecting the day before and the normal EEJ days as the control days, it is intrinsically proved that the regular eastward electric field for the whole day at the equator was not just weakened but actually was flipped for several hours by the influence of tides related to the spectacular Sun-Moon-Earth alignment near the middle of the day. The effect of flipping the electric field was clearly seen in the equatorial ionosonde data and through the large array of GPS receivers that accomplished the TEC data. The main impact of the change in the electric field was the reduced EIA at all latitudes, with the anomaly crest that shifted towards the equator. The equatorial F-region density profile was also showing an enhanced F region peak in spite of a reduced VTEC. By comparison to the plasma density depletion associated with the temporary lack of photo-ionization during the episode of eclipse, the electrodynamical consequences of the eclipse were far more important and influenced the plasma density over a wide range of latitudes.
\end{abstract}


\end{frontmatter}


\section{Introduction}
\label{Introduction}
The dynamics of the equatorial ionosphere is greatly influenced by the enhanced cowling conductivity near the dip equator. The manifestation of this is seen as an intense current, flowing eastward in the ionospheric E-layer during the daytime, within a narrow belt of $\pm3^{\circ}$ latitude over magnetic dip equator, which is called as equatorial electrojet, EEJ \citep{chapman1951}. In the vicinity of the dip equator, it is seen as daytime enhancement in the horizontal component (H) of the geomagnetic field. However, it is also observed that at certain geomagnetic quiet days the EEJ current flows in the reverse (westward) direction, generally during the morning and evening hours, giving rise to negative depression in the H which is termed as counter electrojet, CEJ \citep{Gouin1967}. Any equatorial phenomena causing disturbance in the net zonal electric field eventually affects the E and F-region and hence the distribution of F-layer plasma in the low latitude ionosphere. The total electron content (TEC) measured from the GPS observables is powerful tool to study the static as well as dynamic characteristics of the F-region plasma and its photochemistry, as main contribution to the TEC comes from the F-region. The vertical drift of F region plasma gives rise to Equatorial Ionization Anomaly (EIA) in the low latitude belt. The EIA is characterized by a trough in the ionization density at the geomagnetic equator and crests on either side of the equator within $\pm 15^{\circ}$ magnetic latitudes \protect{\citep{Appleton1946}}. The development of EIA gets affected by various transient processes like geomagnetic storms, prompt penetration electric fields, solar eclipses, etc. The effects associated with eclipses alter the whole electrodynamics of the low latitude ionosphere specially when eclipse occurs near the dip equator. Hence, the underlying physical processes during the eclipses can be probed by studying the variations in EEJ, TEC, and other ionospheric parameters.\\

Solar eclipse has a paramount importance for researchers as it provides a unique opportunity to investigate the atmospheric and ionospheric effects caused by the rapidly moving shadow of the Moon over the Earth, in addition to enhanced gravitational tides. The effects associated with a solar eclipse is always unique, since the solar eclipses may occur at various geographic locations during different phases of the solar activity, geomagnetic conditions, seasons, as well as local times of the day. Moreover, a particular eclipse does not repeat in any recognizable pattern due to the diverse Sun-Moon-Earth alignment geometries and different lunar orbital characteristics \citep{espenak2006five}; hence the study of individual eclipses becomes necessary. Also, induced ionospheric effects have significant impact on satellite-based trans-ionospheric radio-telecommunication and navigation systems \citep{Vyas2012}. The obscuration of solar radiation during partial, annular or total solar eclipse, consequences spatial and temporal ionospheric and thermospheric variations with reduced production of electrons and accelerated recombinations \citep{rishbeth1969introduction}. Ionospheric responses to the earlier solar eclipses have been broadly studied by many researchers across the globe with different techniques, such as incoherent scatter radar \citep{Evans1965,Salah1986}, Faraday rotation measurement \citep{Klobuchar1965}, ionosonde \citep{Adeniyi2007}, ground and space based GPS observations \citep{Tsai1999,Tomas2008}, satellite \citep{Cohen1984}, rocket-borne measurements \citep{Manju2012, Manju2014}, and magnetometers \citep{Sridharan2002,Tomas2008}. Similarly, model calculations of the eclipse effects are also carried out by \citet{Le2010,Lin2012}.\\

Earlier studies demonstrate the reductions in the ionization during the eclipses, based on the comparison of the solar eclipse day with one or more control days, mainly the previous and succeeding days of the eclipse  \citep{Evans1965,Adeniyi2007,Chandra2007,Manju2012,MadhavHaridas2012,SanjayKumar2013}. Some studies also have considered mean variations of five international geomagnetically quiet days (Q-Days) or monthly averages excluding the disturbed days \citep{Sharma20101387,Choudhary2011,Sanjaykumar2012}. \protect {\citet{Sridharan2002}, \citet{St.-Maurice2011}, and \citet{Choudhary2011}} demonstrated the effects of the eclipse induced terminators, by using observations and models which is really thought-provoking. Most of the studies ascribe the reduced TEC to the obvious temporary interruptions of the solar radiation due to the obscuration. However, the equatorial dynamics during the eclipse and control days has to be accounted for reliable and better physical picture. One has to be careful in selecting control days as studies reveal that the frequency of CEJ occurrence is more during geomagnetically quiet days due to the localized changes in the zonal electric field, which is more prevalent during the solar minimum solstice months \protect{\citep{Mayaud1977}}. This could result in ambiguity when the eclipse phenomena are compared with the control or quiet days, without verifying the disturbances in the zonal electric field of these days. Note that the disturbances in the zonal electric field is reflected in the magnetometer EEJ data, therefore some information of the equatorial ionosphere can be derived based on the EEJ strength. Hence, present study is based on due consideration of the EEJ strengths on the eclipse as well as the control days. \\

The eclipse of 15 Jan 2010 gave a unique opportunity to understand the changes in the electrodynamics of the ionosphere during the passage of the Moon\textasciiacute s shadow across the Indian region. The eclipse took place during the solar minimum period in the ascending phase of $24^{th}$ Solar cycle and the geomagnetic condition was very quiet. It happened to be the longest annular eclipse of the $3^{rd}$ Millennium with the maximum duration of annularity 11 min 08 s (its duration would not be exceeded until the year 3043), visible over 333 km wide track covering almost half of the Earth \citep{espenak2006five}. The annularity was clearly visible from South Kerala and South Tamil Nadu in India, but other parts of the Indian region witnessed only a partial eclipse. The major attraction of the present solar eclipse is that it occurred during the early afternoon hours, i.e., peak ionization time over the Indian EIA region, and path of the annularity crossed the dip equator over the peninsular part of India. \\

Though there are a number of studies on the occurrence of CEJ during quiet days, very few investigations tried to understand the physical mechanism of CEJs occurring during solar eclipses \protect {\citep{Tomas2008}}. Therefore, it still remains as a controversial issue and needs to understand, whether the CEJ is induced/modulated by the effects associated with reduced radiation during the eclipse or it is due to the  enhanced lunar tidal forces. Also, it could be a superposition of both the effects. We believe that adopting different criteria for the selection of control days gives rise to variations in the inferences drawn from the studies. The reason for selecting this particular solar eclipse is that, it crossed the dip equator during the peak hours of the day when the anomaly development suppose to be at its progressive state. Therefore, in the present work, we set our main objectives of the eclipse study, as (1) to probe the underlying mechanism of the CEJ occurred during the eclipse, (2) investigate the influence of prevailing electrodynamics on the behavior of ionospheric electron densities, and (3) examine latitudinal variations of TEC to access the possible impact of anomalous electrodynamics on the fountain effect and hence the EIA. The availability of global geomagnetic and interplanetary parameters, local magnetometer, ionosonde, and the large array of GPS receivers gave us an opportunity to investigate these objectives with a multi-instrumental study. Above all, the comprehensive explanations of \protect{\citet{Rastogi19899,Choudhary2011,St.-Maurice2011,MadhavHaridas2012,Manju2012, Manju2014}}, motivated us for looking into the peculiar aspects of the exceptional eclipse configuration during the period.

\section{Data used and method of treatment}
The locations of magnetometer (blue square), ionosonde (black square), and GPS (red dot) observation stations used in this study, are marked on the map showing path of the solar eclipse in {\cref{fig:eclipsepath1}} (modified from \citet{eclipsepathEspenak}). The corresponding geographic and geomagnetic coordinates, as well as dip angle of the locations, are listed in {\cref{tab:Allstations}}. Onset time of the eclipse, time of centrality, and the end time of the eclipse, along with the obscuration magnitudes at the observation stations are shown in {\cref{tab:circumstances}}. Note that annularity was observed at Trivandrum and Tirunelveli, whereas at Jaipur the obscuration magnitude was below 50\%, the lowest among all stations in this study. The brief outline of the data and methodology used in our study are set out below. The time used throughout the study is in Indian Standard Time (IST= UT+5:30), as indicated in the figures/tables.

\subsection{\textbf{Interplanetary and Geomagnetic parameters}}
To perceive the interplanetary and geomagnetic conditions during the eclipse period, we  examined the one minute averaged SYM-H (equivalent to hourly Dst values), vertical component of interplanetary magnetic field (IMF-Bz), interplanetary electric field (IEF), and  the 3-hourly Kp index. The SYM-H represents the ring current strength, thereby acts as a proxy for the strength of the magnetospheric disturbance. \\

As described in the introduction, the magnetometer H variation over the equator is related to the $East \rightleftharpoons West$ ionospheric electric current, and hence is a proxy for the strength of EEJ at the equator. Tirunelveli (TIR) magnetic observatory is in the vicinity of EEJ being located close to the dip equator, with EEJ effects superimposed on the worldwide solar quiet (Sq) field, whereas Alibag (ABG) magnetic observatory is an off-equatorial low latitude station, mainly influenced by the global Sq current system. In the present study, the EEJ is calculated by subtracting nighttime baseline H from the daytime value and then finding the difference between the two stations, to ensure the removal of magnetospheric and Sq contributions. Therefore, the EEJ strength is defined as:
\protect{\begin{equation}
EEJ ~strength = \Delta H_{TIR}-\Delta H_{ABG}
\end{equation}}
where, $\Delta H_{TIR}$ and $\Delta H_{ABG}$ are magnetic field variations at TIR and ABG respectively. This method was earlier suggested by \citet{Chandra1971} and later used by several workers \citep{RastogiandKlobuchar, Manoj2006, Sripathi2012, Bhaskar2013}. The present study uses this EEJ strength to investigate the electrodynamics of the equatorial ionosphere during the eclipse and the control days.

\subsection{\textbf{Ionosonde data}}
The F region parameters (10 minute resolution) are measured by two different equatorial ionosonde stations, Trivandrum (Digisonde Portable Sounder 4D; 0.5-30 MHz) and Tirunelveli (Canadian Advanced Digital Ionosonde; 2-20 MHz). The importance of simultaneous consideration of these two stations is that, they are situated (1) on the equatorial belt and (2) in the umbral shadow zone of the annular solar eclipse. Thiswould help in avoiding any malfunctions in the instruments and thereby an unambiguous understanding of the eclipse-induced changes in (a) equatorial electrodynamics governed by $E \times B$ drift and diffusion, and (b) ionospheric processes (such as recombination loss, reduced production etc). F-region electron density is the major contributor to the ionospheric TEC derived from the GPS observables and other techniques, hence detail inspection of the F region parameters (foF1, foF2, and hmF2) is carried out in the present study.  \\

 \subsection{\textbf{GPS derived TEC}}
Continuous operation of a network of GPS receivers offers an excellent opportunity for exploring spatial and temporal variations of TEC over the globe. The TEC is quantified by counting the number of electrons in a vertical column with a cross sectional area of $1 m^{2}$ and is measured in a unit called TECU, where $1 ~TECU = 1\ast10^{16} electrons/m^{2}$ \citep{Klobuchar1991}. In this study, we examine the behavior of ionospheric TEC derived from  dual frequency GPS data in the Indian region at Tirunelveli, Kodaikanal, Bangalore, Hyderabad, Vizag, Mumbai, Shillong, and Jaipur during the eclipse and control days (geophysical parameters are listed in \cref{tab:Allstations}). The GPS stations at Bangalore and Hyderabad are under International GNSS Service (IGS) network. Recorded data at Tirunelveli, Vizag, Mumbai, Shillong, and Jaipur GPS stations are obtained from the Indian Institute of Geomagnetism, Navi Mumbai, India. The GPS station at Kodaikanal is operated by the Indian Institute of Astrophysics, India. Observation data from all these stations are processed using the GPS-TEC analysis application software, developed by Gopi Seemala \citep{Seemala2011}. The software algorithm calculates line-of-site TEC, i.e., the slant TEC (STEC) along the ray path, using the differential phase advance and group delay measurements of GPS L1 (1575.42 MHz) and L2 (1227.60 MHz) frequencies. The TEC corresponding to group delay can be written as:
 
 \begin{equation}
 TEC_{group}=\frac{1}{40.3}\left( \frac{f_{1}^{2}f_{2}^{2}}{f_{1}^{2}-f_{2}^{2}}\right)\left( P2-P1\right) 
 \end{equation}
 here, $P1$ and $P2$ are the pseudoranges (groups) corresponding to GPS frequencies $f1$ and $f2$.
 Similarly, the relative phase TEC can be estimated from the phase measurements on both the frequencies as:
 \begin{equation}
 TEC_{phase}=\frac{1}{40.3}\left( \frac{f_{1}^{2}f_{2}^{2}}{f_{1}^{2}-f_{2}^{2}}\right)\left(\Phi1-\Phi2\right) 
 \end{equation}
 where, $\Phi1$ and $\Phi2$ are phases of carrier frequencies $f1$ and $f2$ respectively. \\ 
 
 The TEC derived from differential group delays have large uncertainties due to high noise, hence it is smoothed by carrier phase leveling. However, the estimated TEC from the above method is actually the STEC which is then converted into its vertical equivalent (VTEC), using a single layer model mapping function \citep{Schaer1999}. To minimize the changes in satellite geometry, effect of multipath and other lower atmospheric attenuations, an elevation angle of $30^{\circ}$ at 350 km altitude is used while getting VTEC from STEC at the ionospheric pierce points (IPPs).\\

 \section{Observations}
 \label{Observations}
 \subsection{\textbf{Interplanetary and Geomagnetic parameters}}
 \cref{fig:geoindices1} shows one minute averaged interplanetary electric and magnetic fields (IEF and IMF-Bz), SYM-H index, and 3-hourly Kp index during 15 Jan 2010 eclipse day. It can be observed from the figure that the $\Sigma Kp$ index was 8 and maximum negative excursion of SYM-H index was around -14 nT. The IMF-Bz had a small magnitude ranging between -5 to 5 nT. Similarly, the variation of the electric field is seen to be ranging from -2 to 2 mV/m. Although the southward and northward flipping of IMF-Bz and corresponding duskward and dawnward IEF (-V X Bz) occurs during the initial phase of the eclipse, it is very small to build up any notable magnetospheric disturbance. This suggests that the eclipse day was geomagnetically quiet, having a very small influence of interplanetary disturbance. 
 
 \subsection{\textbf{EEJ Strength}}
 \label{eejstrength}
 {\cref{fig:eejstrengths}(a) shows hourly EEJ strength for the eclipse and its adjacent days, as well as 5 Q-Days of the month. It is clear from the figure that the EEJ was considerably weakened or reversed for three consecutive days, with the largest effect being witnessed on the eclipse day.} In particular, the eclipse day itself begins with a moderate eastward electrojet until 09:30 IST, followed by a substantial decrease, resulting in a strong CEJ around 14:30 IST (about 1 hr after maximum phase of the eclipse) and later recovers to its normal strength. Therefore, the whole period of this eclipse falls under the influence of strong CEJ. As noted earlier in the Introduction section, selection of appropriate control days is an imperative aspect to draw realistic inferences from the observations. In connection to this, we studied the diurnal EEJ variation for 2010 year and found maximum occurrences of suppressed EEJ strength or CEJ during the month of January. Therefore, we tried to choose the most suitable reference/control days on the basis of their EEJ strength and normalcy in pattern. We examined the 10 Q-Days of the January month and interestingly observed that most of them are also perceived suppressed EEJ or CEJ effects, either in the morning or afternoon hours. Moreover, by considering the mean of 5 Q-Days as shown in \protect{\cref{fig:eejstrengths}}(b), the EEJ strength shows reduced intensity during the morning hours, and there is a shift in the EEJ peak from the regular noon-time, apparently due to the cumulative effect of suppressed EEJ strength on most of the Q-Days. From the above observations, we realized that neither the adjacent control days nor the mean Q-Days can solely consequence the complete picture of what actually happened on the eclipse day. Therefore, we selected two control days; (1) the preceding day (14 Jan 2010) manifesting an analogous pattern of CEJ but with a relatively weaker strength as compared to the eclipse day and (2) a very normal EEJ day (2 Jan 2010) with the regular EEJ strength and pattern, i.e., peaking near noon and having the positive value.\\

 \subsection{\textbf{Variations of F region parameters}}
 \label{F-RegionParameters}
 \cref{fig:ionosonde parameters} depicts  the F-region parameters (foF1, foF2, and hmF2) at equatorial stations (Trivandrum and Tirunelveli) and variation in the EEJ strength (bottom panel) on the eclipse day (ED: 15 Jan 2010) and the control days (CD1: 2 Jan 2010; CD2: 14 Jan 2010). A clear depletion in foF1 from 4.5 MHz to $\sim$3 MHz is markedly noticed in {\cref{fig:ionosonde parameters}} (top panel), during the eclipse period at both the stations. However, the depletion started with a delay of about 20 minutes from the onset of the eclipse. It is also clear from the figure that during the normal EEJ day (CD1), the foF2 increases after sunrise, attains the pre-noon peak around 09:00 IST, followed by a second peak after sunset. The regular day-time minimum of foF2 ($\sim 5.5$ MHz) observed at both the stations corresponds to the maximum EEJ strength over the equator. On the eclipse day, the pre-noon foF2 (above 7.5 MHz) is higher than CD1 ($\sim 6.5$ MHz) much before the start of the eclipse, whereas the difference between the eclipse and its preceding day, CD2 is $ < $1 MHz at both the equatorial stations. However, in the afternoon lesser foF2 is observed on the eclipse day. Although the second peak ($\sim 7.5$ MHz) is observed at around 18:00 IST, unlike 2 Jan 2010, it did not settle down later. Further observations of hmF2, presented in the \cref{fig:ionosonde parameters} (second panel from the bottom), shows that on the eclipse day, after having a normal increase until 09:30 IST, it started lowering abnormally over a period of about 2 hrs, with a series of fluctuations. However, after 12:00 IST, significantly sharp upward and then downward transitions in hmF2 is noticed at both the ionosonde stations. Note that, hmF2 unusually remained at the lower altitude until about 16:00 IST, and later increased slightly during the sunset hours. \\

\subsection{\textbf{Variations of TEC}}
To complement the ionosonde observations at the magnetic equator, we considered the GPS derived VTEC data at the Tirunelveli station. The 15-min averaged VTEC variations along with obscuration magnitudes (top) and the EEJ strengths (bottom), are plotted in {\cref{fig:TECvariation_Tirunelveli}}. The figure clearly shows the noontime bite out with two peaks (pre-noon and afternoon) in VTEC on the normal EEJ control day (CD1). However, overall substantial increase in VTEC is noticed on the day before (CD2) of the eclipse day, which corresponds to the maximum CEJ in the afternoon hours. Unlikely, on the eclipse day, it is interesting to see the depletion in regular VTEC with the onset of CEJ around 09:30 IST, well before the start of the eclipse while the ionosonde data evidences an anomalous increase in foF2 as mentioned in \protect{\cref{F-RegionParameters}}. After the recovery of CEJ, following the maximum phase of the eclipse, it took a long time to achieve the normal VTEC level at this stations. Again to analyze the eclipse accompanied anomalous CEJ effects beyond the magnetic equator, the temporal variations of VTEC at different latitudes across the Indian region are verified in {\cref{fig:AllstationTEC}}. The respective obscuration magnitudes, start, maximum, and end time of the eclipse, and the departures of VTEC from the two control days are provided in \protect{\cref{tab:circumstances}}. The normal EEJ day VTEC variation depicts two diurnal peaks at the magnetic equator, gradually merging towards the higher latitudes to represent a single diurnal peak, but is absent on the eclipse day. An important aspect of the preceding control day TEC around the equator is that its value is higher than the eclipse and the normal EEJ control day. Whereas, at higher latitudes, normal EEJ control day shows greater TEC value than that observed during the eclipse and preceding control days.\\

 It is evident from {\cref{fig:AllstationTEC}} that the VTEC on the eclipse day was markedly lowered at all stations w.r.t the control days. The lowering of the regular VTEC started after 10:00 IST at all stations with maximum excursions during the recovery or the post eclipse period. The eclipse day deviations from the preceding control day appear to be slightly lesser than the deviations from the normal EEJ control day. However, the deviations are more around latitudes of Vizag which falls in the vicinity of regular EIA crest (around {$10^{\circ}$} during the season) and could not recover its normal VTEC level after end of the eclipse. Interestingly, a clear pre-eclipse enhancement of TEC is noticed at this station w.r.t both the control days in the off-equatorial stations. The low value of TEC at Shillong and Jaipur during the control days suggests that their latitudes lay beyond the EIA crest during the January month which commensurate with the past studies of \protect{\citet{Bhuyan2009}}, illustrating the general behavior of low EEJ strength during the winter solstice and the eclipse falling within a solar quiet epoch. However, the obscuration magnitude at Shillong was more than at Hyderabad and Mumbai. The onset of the eclipse at Shillong was observed about one hour later as compared to the equator-ward stations due to North-Eastward progress of the eclipse shadow. Exceptionally, the temporal VTEC at Jaipur was down on the eclipse day right from morning to the end, though further depletion was perceived after about 9:30 IST and could not recover to normal level even by 20:00 IST. This was quite anomalous and needs further investigation. In brief, the lowest deviation in VTEC from the normal EEJ control day is noticed at Tirunelveli, followed by intermediate deviations at equator-ward low latitude stations (Kodaikanal and Bangalore), and highest departure near latitudes of Vizag and Hyderabad (regular crest latitude). \\

In \protect{\cref{tab:circumstances}}, we compare the percentage deviation of VTEC at all the studied stations and the corresponding deviations reported by earlier researchers who selected different quiet/control days as reference days for the same event. It is evident from the tabular values that, the estimated \% deviations in VTEC with dual control days reasonably differs from past reports at the same or nearby locations in the same eclipse event. \protect{\citet{Vyas2012}} though report reductions in VTEC at Trivandrum (equivalent to our Tirunelveli station), Bangalore, Hyderabad, Kolkata and Guwahati (equivalent to our Shillong and Jaipur stations) by comparing with the mean quiet period (9-20 Jan 2010), the reductions are less than our values. \protect{\citet{Galav2010}}, \protect{\citet{Sanjaykumar2012}}, and \protect{\citet{MadhavHaridas2012}} also reported TEC reductions during the same eclipse. They estimated the reductions at different locations by comparing the eclipse day TEC with the averages of three preceding days (11-13 January 2010), the mean quiet days, and the successive control day (16 Jan 2010) respectively. There exists a discrepancy between their reports and our results on TEC reduction which is due to the difference in selection of control days. To the best of our knowledge, there are no earlier TEC studies of this event at Kodaikanal and Mumbai. Nevertheless, \protect{\citet{Momani2011}} reported substantial decrease in TEC at Guangzhou and Hainan (China) and at Okinawa (Japan), in the East Asian longitudes, by comparing the eclipse day VTEC with respect to the previous, successive, and averaged Q-Days of the month whose likeness is reflected in \protect{\cref{tab:circumstances}}.\\

To examine the anomaly strength and its latitudinal extent on the eclipse and the control days, hourly averaged VTEC values for the period 09:00\textendash17:00 IST, are plotted against geomagnetic latitudes of the GPS stations in {\cref{fig:LatitudinalTEC}}. The figure clearly depicts that on the normal EEJ control day (2 Jan 2010), the EIA crest is distinctly visible near Vizag (Geomag. latitude: 8.50 N) around 14:00 IST. However, on the eclipse as well as the preceding day, the EIA crest is observed to be shifted to lower latitude, near Bangalore (Geomag. latitude: 4.29 N) at about 13:00 IST. It is intriguing to see that, though the overall TEC was lesser on the eclipse day than the day before, the anomaly crest developed around identical latitudes. {\cref{fig:PercentagedeviationofTEC}} shows latitudinal plots of percentage deviation of TEC from the preceding day {\cref{fig:PercentagedeviationofTEC}}(a) and normal EEJ control day {\cref{fig:PercentagedeviationofTEC}}(b), during pre- and post- eclipse as well as the intermediate phases (initial, main, and recovery) of the eclipse. The pre-eclipse \% deviation of TEC is enhanced near Hyderabad, when compared with the preceding control day whereas, when estimated with respect to the normal EEJ control day, crests are seen at Kodaikanal and Mumbai. Nevertheless, the reduction started at all the stations right from the initial phase, with the maximum excursion being witnessed during the post-eclipse period. The other notable feature is that the \% deviations at different phases could be distinguished when the deviations are evaluated with reference to the previous day, while the normal EEJ control day is considered as reference day, the different phases are mixed up.\\

\section{Discussion}
The first noticeable feature in the magnetometer observations in {\cref{fig:eejstrengths}} is that the EEJ strength was considerably weakened or reversed for a period of 3 consecutive days, with highest CEJ intensity being witnessed on the eclipse day, in spite of magnetically quiet conditions. This indicates the reversal of the regular daytime electric field. For the same eclipse event, clear reversal in both the regular westward zonal wind and the poleward meridional wind, were confirmed by \protect{\citet{Manju2012}}, through the rocket-borne in situ Electron density and Neutral Wind probe (ENWi) measurements above 100 km. This indicates that the eastward zonal wind reversed the polarization electric field which manifested as a strong CEJ. Similarly, observations from the ionosonde at Trivandrum and Gadanki in India confirm that the meridional wind direction was poleward in the pre-eclipse period (shadow south of equator), equatorward during the eclipse main phase (shadow over the equator), and again poleward in the post-eclipse period (shadow north of Gadanki) in the northern hemisphere \protect{\citep{MadhavHaridas2012}}.\\

The presence of westward electric field results in downward plasma motion which is pronounced on the eclipse day. Typically, the variation of plasma density extending from E to F1 region is believed to be due to photo-ionization production and recombination losses, whilst the F2-region variations, apart from these two processes, are affected by temperature variations, horizontal and vertical transport and photochemical reactions \citep{Rishbeth1968}. However, the observed time lag (about 20 min) between the eclipse onset and response of the F1-layer might be due to the short lifetime of the electrons in this layer. The lifetime of the F1 region electrons is $\sim 1000$ seconds around altitudes of 200 km \protect{\citep{banks1973}}. These suggest that the lowered photo-ionization production due to the eclipse obscuration and continued recombination, disturbed the equilibrium at the foF1 region. Adversely, the F2-region shows a clear signature on eclipse day and a subtle difference from the usual on the day before, with increase in foF2 and decrease in hmF2 that corresponds to the faster downward motion of plasma at the equator. All these variations are also in good agreement with earlier F-region studies by \protect{\citet{St.-Maurice2011}, \citet{Nayak2012}, and \citet{Manju2014}}. Typically, the dynamics of the daytime equatorial ionosphere is greatly influenced by the eastward zonal E-region electric field. This zonal electric field generates a vertical drift of F-region plasma which in turn diffuses along the geomagnetic field flux tubes towards higher latitudes on both sides of the magnetic equator to develop the EIA. Hence, smaller amplitude of diurnal variation of plasma density as well as VTEC are expected over the equator. However, on the eclipse day, weakening of the regular eastward zonal electric field and the temporary interruptions in the solar illumination completely altered the dynamics and chemistry of the equatorial ionosphere.\\

As illustrated in the \cref{F-RegionParameters}, the overall increase in foF2 and decrease in hmF2 level in the equatorial ionosondes, are predominantly attributed to faster downward movement of the F region plasma on the eclipse day while far stronger CEJ was at its progressive state. Remarkable explanation has been given by \protect{\citet{Choudhary2011}} in support of the above situation, through running a 1-D model, which includes the associated dynamics and chemistry. Their model results realized that significant portion of the F-layer plasma was actually moving downwards after 10:00 IST which later suffered an abrupt upward motion in the early afternoon, shortly before the eclipse maximum, followed by sharp downward movement afterwards. In particular, the abnormal electrodynamics introduced by the eclipse induced neutral wind pattern and the E-region conductivity gradients, was the primary driving force, pushing the plasma downwards for an extended period whilst chemical recombination was more at the lower altitudes.\\

Aside from the model derived F region drift between 10:00-12:00 IST which indicates the westward zonal electric field, the EEJ data as obtained by \protect{\citet{Choudhary2011}} for the same period, indicated an eastward electric field (though the magnitude was apparently low) which was quite unusual and remained unexplained in their analysis. We believe that the disagreement of their EEJ value and the model drift could be due to their different approach in calculating the CEJ from the equatorial Tirunelveli magnetometer H data, in which they used SYM-H index instead of Alibag (off-equatorial) magnetometer H data, for removal of the global magnetospheric current contribution. They didn't follow \protect {\citet{RastogiandKlobuchar}} method, presuming that during the solar eclipse the relationship between perturbation of Sq and EEJ could deviate from the usual. However, we have used the usual method of calculating the EEJ/CEJ by subtracting the nighttime baseline H from the daytime value and then finding the difference between Tirunelveli and Alibag stations, which indicated an excellent agreement with their deduced drift during the period. This suggests there is hardly any divergence in the relationship between perturbation of Sq and EEJ during the eclipse. In connection with this, \protect{\citet{St.-Maurice2011}} investigated the electric field development driven by the interactions between the low (eclipse) and the high pressure (solar heating) systems. They clearly demonstrated that the already weakened EEJ after 9:00 IST was further reduced by the low-pressure system, associated with the eclipse over the equator, to prompt a strong CEJ in the afternoon hours.\\

The sharp upward and downward oscillation of hmF2 during the eclipse main phase, is probably due to the accumulation of negative charges near the eclipse terminators, which is similar to the mechanism of the Pre-reversal-enhancement (PRE) proposed by \protect{\citet{Farley1986}}. The probable simultaneous occurrences of the eclipse induced and the regular evening PREs were earlier proposed by \protect{\citet{Sridharan2002}}, during the dusk-time total solar eclipse (11 Aug 1999) over the equator. Later, the eclipse induced PRE was confirmed by \protect{\citet{St.-Maurice2011}} for the present event, through the formalism based on the superposition of the low (cooling) and the high pressure (solar heating) wind systems generated during eclipse. Subsequently, \protect{\citet{Choudhary2011}} have supported the eclipse PRE signature through effectively retrieving the direction and amplitude of the plasma drift over the equator by constructing the numerical 1-D model. As an alternative explanation, some earlier reports describe the sharp uplifting of hmF2 due to the rapid decrease in electron density at lower F-region, uplifting the normal F2 layer to higher altitude. After eclipse maximum passed, the ionization in the lower altitude progressively recovers and develops a new layer, with gradual shrinking of the upper layer which perceives as a downward movement of hmF2 \protect{\citep{Liu2000,Adeniyi2007}}.\\

The relatively lower equatorial VTEC on the eclipse day from the day before, in spite of far stronger excursion in the zonal electric field, is clearly anomalous. Here, it should be noted that, the major portion of the regular VTEC corresponds to the F2 region electron density in the ionosphere. However, although the foF2 was higher during the eclipse day, the lower altitude regions suffer larger reduction in density due to the eclipse induced increased chemical recombinations that happened to be the main convict for the decreased columnar content (local VTEC). All these observations agree well with the model calculations and the inferences drawn in the earlier studies \protect{\citep{Choudhary2011, St.-Maurice2011,MadhavHaridas2012,Manju2012}}. However, akin behaviour of the foF2 and TEC experiencing reductions after the peak eclipse phase is due to the inhibited electrodynamics, increased poleward wind, and failure in recovering at the lower altitudes in the absence of solar ionization as the Sun had already moved westward \protect{\citep{MadhavHaridas2012}}. As observed from {\cref{fig:AllstationTEC}}, the markedly lowered VTEC at all the stations on the eclipse day, relative to the control days, are predominantly attributable to the weakened EIA and lesser local ionization. The reversal of the ionospheric zonal electric field at the equator forced large portion of the lower F-region plasma downward during most of the afternoon hours that indicates a weakened fountain effect and poor feeding of the EIA. However, some plasma could always makes it to high enough altitudes (400 km and more) which get caught by the downward pull along the magnetic field lines to develop the crest at latitudes close to the equator \protect{\citep{Choudhary2011}. Favorably, in \cref{fig:LatitudinalTEC} we observed that on the day before as well as the eclipse day, the EIA crest was formed near Bangalore (close to equator), rather than the regular development around Vizag. Moreover, the local degree of obscuration has a progressive role towards higher latitudes (deprived of diffusing plasma through fountain effect) though its effect is negligible in the circumstances of the driving equatorial electrodynamics.\\

The other important observation is that the \% deviation of eclipse day VTEC at the equatorial/off-equatorial stations found to be greater/lesser than the preceding day (CEJ day) as opposed to the normal EEJ day. The pre-eclipse VTEC enhancement is clearly evident on the eclipse day when normal EEJ day was used as reference. However, enchantment is less prominent when estimated with reference to the preceding day which could be due to the presence of CEJ on the preceding day. As concerns, selecting two different types of control days facilitated us for a clear understanding of prevailing equatorial electrodynamics and its effects at the equator and elsewhere, during the eclipse. We presume the maximum negative deviations in VTEC at Vizag due to the weakened EIA, with the crest moved towards the equator from the regular development near Vizag. Marked differences are noticed between the \% deviation of the eclipse day TEC and the earlier reports of \protect{\citet{Galav2010,Momani2011,Sanjaykumar2012,Vyas2012,MadhavHaridas2012}}. In light of these discrepancies from earlier reports, it is evident that the selection of an appropriate control day is very much crucial. This is particularly true in the equatorial and low latitude region during the solar eclipse. Moreover, considering the normal EEJ day helped us to estimate the net ionospheric effects on the eclipse day, whereas preceding CEJ day supported us to elucidate the prevailing role of large scale electrodynamics around the eclipse day.\\

Although the modified ionospheric electrodynamics during the eclipse is predominantly attributed to the weaken/reversed daytime electric field manifesting CEJ in the magnetometer data, the possible origin of the CEJ has not been discussed in the earlier studies. As noted earlier, the preceding and the succeeding days of the eclipse day, also experienced pretty good CEJ but with a relatively lesser amplitude than the eclipse day. Therefore, we suggest that the strong gravitational tidal effect does exist on the solar eclipse day, plausibly due to the exceptional Sun-Moon-Earth alignment. Note that, during a solar eclipse, the solar (S) and lunar (L) gravitational tides should be in phase and thus one can expect the strong semi-diurnal tidal forcing in the dynamo region of the ionosphere \protect{\citep{Yamazaki2012}. Additionally, the occasional temperature fluctuations in the lower atmosphere, mostly in the December and January months, may also multiply the amplitude of the geomagnetic luni-solar tides, even the lunar tides become larger than the geomagnetic solar tides \protect{\citep{Yamazaki2012}}. Moreover, there are also evidences of exceptional global enhancements in geomagnetic lunar tides during the January month due to the equatorial lunar effects \protect{\citep{Mayaud1977,Stening1977}. All these situations may modulate the daytime eastward electric field which could be manifested as CEJ in the magnetometer recordings. Furthermore, the high pressure in the southern hemisphere during the winter and the low pressure along the eclipse path also likely to be intensifying/generating the CEJ. However, the effect of pressure difference has a subordinate role as compared to the accompanying spectacular gravitational tides. Favorable existence of CEJ around the New and Full Moon are endorsed by \protect{\citet{Rastogi19899}} and \protect{\citet{Tomas2008}}, indicating their lunar tidal dependences. Since solar eclipse occurs during New Moon, CEJ formation is favored, particularly for daytime passage of the lunar shadow over the dip equator. From the evidences of frequent CEJs during the January and, above all, the far stronger CEJ around the eclipse day, we stress the tidal forces are the major culprit for what happen on the day. Apart from these, the frequent fluctuations in the ionospheric parameters are most likely to be due to the atmospheric gravity waves, generated from the thermal instability during the eclipse. The existence of these gravity waves during the event are though confirmed earlier by  \protect{\citet{Dutta2011}, \protect{\citet{Sumod2011}}, and \protect{\citet{Manju2014}}, the aspect is out of context in our study.\\

\section{Summary and conclusions}
In summary, we can ascertain that the faster downward movement of plasma was the primary cause for increased electron density over the equator and substantially decreased TEC at all latitudes, whilst temporary lack of solar radiation had a subdued role. Although the decrease in foF1 is clearly related to the eclipse obscuration as expected, the increase in foF2 and lowering of hmF2 are the major consequences of the downward forcing of plasma around the eclipse. The largest deviations in VTEC from the normal day, at the regular crest latitude, is due to the equator-ward shifting of the anomaly crest. This is attributable to the severely weakened fountain effect and poor feeding of the EIA. The potential driver behind the downward plasma motion on the eclipse day was the far stronger westward zonal electric field. This is probably due to the exceptional electrodynamics resulted through the combined modulated effects of (1) enhanced tide, (2) gravity waves, and (3) the low pressure wind system setup during the eclipse day. However, the existence of considerably weakened or reversed EEJ for 3 consecutive days, with largest effects on the eclipse day, substantiate the abnormal modulation of the gravitational tides, which is likely to be due to the incredible Sun-Moon-Earth alignment during the middle of the day.\\

The present solar eclipse gave us a unique opportunity to study the large-scale effects of the altered electrodynamics over the equatorial and low latitude Indian region. The consequences of this study would certainly help in better understanding of various solar-terrestrial phenomena and eventually in establishing more realistic regional ionospheric models over the low latitudes.\\

 \section*{Acknowledgments}
The authors acknowledge CDDIS for IGS-GPS observation data and broadcast ephemeris files, Director, Indian Institute of Geomagnetism for providing magnetometer data, ionosonde data of Tirunelveli and GPS observation data of permanent GPS stations; and B. C. Bhatt from Indian Institute of Astrophysics for providing Kodaikanal GPS data. The ionosonde data of Trivandrum station is obtained from T. K. Pant, Vikram Sarabhai Space Centre (VSSC), Trivandrum, India. Also, sincere appreciation goes to World Data Center for Geomagnetism, Kyoto for information on geomagnetic indices. The OMNI data (1-min averaged SYM-H, IMF-Bz, and Electric field) were obtained from the GSFC/SPDF OMNIWeb interface at \url{http://omniweb.gsfc.nasa.gov}. Three-hourly Kp index is obtained from Space Physics Interactive Data Resource (SPIDR). The authors also thank CODE-University of Bern for GPS satellite inter-frequency biases. The authors also sincerely thank Gopi K. Seemala, Indian Institute of Geomagnetism (previously at Kyoto University Japan) for providing the GPS TEC analysis software used in this study. Our sincere thanks to A. Nishida and Y. Yamazaki for valuable discussions.

 \newpage \noindent
 
 \bibliographystyle{elsarticle-harv} 
 \bibliography{database3}

 
 
 
 


 \begin{figure}[htb]
 \centering
 \includegraphics[angle=180,width=6.5in]{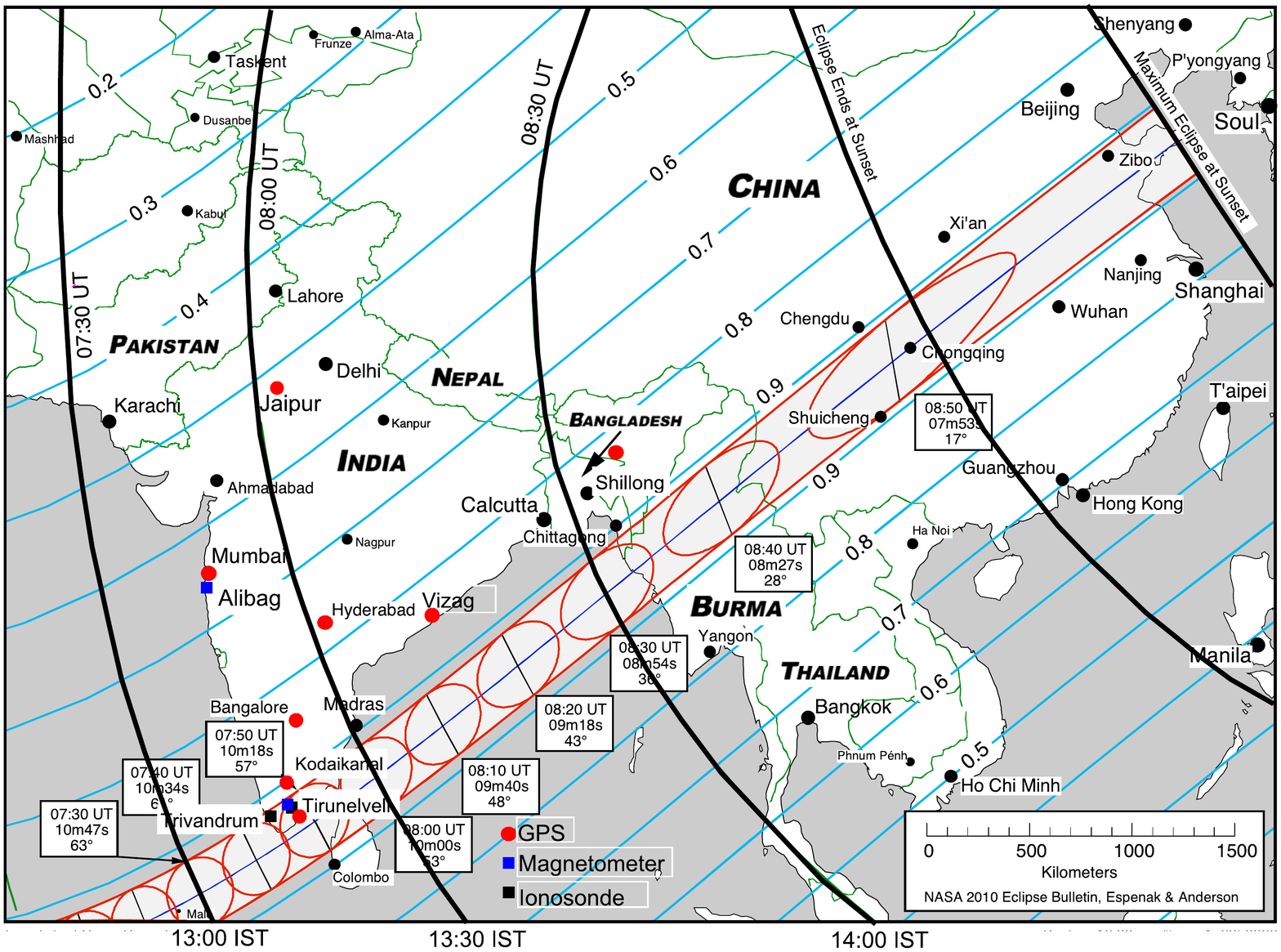}
 \caption{Locations of Magnetometer (blue square), Ionosonde (red square), and GPS receiver (red dot) stations and path of the annular solar eclipse of 15 Jan 2010, modified from \citet{eclipsepathEspenak}.}
 \label{fig:eclipsepath1}
 \end{figure}

\begin{table}[b]
\caption{Geographic and Geomagnetic co-ordinates and Dip angle at observation stations and respective instruments.}
\label{tab:Allstations}
\centering
\begin{tabular}{lcccccc}
\hline 
\rule[0ex]{0pt}{3ex} Station Name & \multicolumn{2}{c}{Geographic}  & \multicolumn{2}{c}{Geomagnetic}  & Dip & Observations\\
\cline{2-5}
\rule[-2ex]{0pt}{0ex} & Lat.& Long.& Lat.& Long. & Angle & used\\
\hline
\rule[-2ex]{0pt}{5ex} Trivandrum (TM308) & 08.54 & 76.87 & 0.11S & 149.46 & 2.36 & ION\\ 
\rule[-2ex]{0pt}{2ex} Tirunelveli (TIR) & 08.70 & 77.80 & 0.03S & 150.39 & 2.67 & MAG, ION, GPS\\ 
\rule[-2ex]{0pt}{2ex} Kodaikanal (KODI) & 10.23 & 77.46 & 01.52N & 150.20 & 6.44 & GPS\\ 
\rule[-2ex]{0pt}{2ex} Bangalore (BAN2) & 13.03 & 77.51 & 04.29N & 150.49 & 13.14 & GPS\\ 
\rule[-2ex]{0pt}{2ex} Hyderabad (HYDE) & 17.42 & 78.55 & 08.55N & 151.87 & 23.03 & GPS\\
\rule[-2ex]{0pt}{2ex} Vizag (VSKP) & 17.74 & 83.33 & 08.50N & 156.49 & 23.43 & GPS\\  
\rule[-2ex]{0pt}{2ex} Alibag (ABG) & 18.62 & 72.87 & 10.24N & 146.53 & 25.94 & MAG\\
\rule[-2ex]{0pt}{2ex} Mumbai (PANV) & 19.01 & 73.10 & 10.60N & 146.80 & 26.74 & GPS\\  
\rule[-2ex]{0pt}{2ex} Shillong (SHIL) & 21.56 & 91.86 & 15.79N & 165.07 & 31.1 & GPS\\
\rule[-2ex]{0pt}{2ex} Jaipur (JAIP) & 26.87 & 75.82 & 18.14N & 150.16 & 41.39 & GPS\\
\hline
\multicolumn{7}{l}{ION-Ionosonde, MAG-Magnetometer, GPS-Global Positioning System}
\end{tabular}
\end{table}

\begin{landscape}
 \begin{table}[htb]
 \caption{Local circumstances and percentage deviation of TEC during annular solar eclipse of 15 Jan 2010 from the normal EEJ control day (CD1; 2 Jan 2010) and preceding control day (CD2; 14 Jan 2010) at different locations and their comparisons with the earlier reports.}
 \label{tab:circumstances}
 \centering
 \begin{tabular}{cccccccc}
 \hline \rule[0ex]{0pt}{3ex} Station & Start & Maximum  &  End  & Obscuration &  \multicolumn{2}{c} {$\Delta \%$ TEC($\downarrow$)(Our Observations)} &  $\Delta \%$TEC($\downarrow$)(Others)  \\ 
 \cline{6-7}
 \rule[0ex]{0pt}{3ex}&&&&& 2 Jan 2010(CD1) &	14 Jan 2010(CD2)\\
 \hline 
 \rule[-2ex]{0pt}{5.5ex} Trivandrum & 11:04 & 13:14 & 15:05 & 0.844 &  NA & NA & 10-15* \\ 
 \rule[-2ex]{0pt}{3ex} Tirunelveli & 11:08 & 13:17 & 15:06 & 0.844 & 25 & 31& NA \\ 
 \rule[-2ex]{0pt}{3ex} Kodaikanal & 11:10 & 13:18 & 15:08 & 0.833 & 30 & 32& NA \\ 
 \rule[-2ex]{0pt}{3ex} Bangalore & 11:16 & 13:23 & 15:11 & 0.772 & 39 & 35& 20\textmusicalnote, 15-30*, 40\dag, 30\ddag\\ 
 \rule[-2ex]{0pt}{3ex} Vizag & 11:44 & 13:43 & 15:22 & 0.778 & 56 & 42& 20-35* \\
 \rule[-2ex]{0pt}{3ex} Hyderabad & 11:29 & 13:32 & 15:15 & 0.688 & 48 & 18& 25-35*,32-36\dag \\ 
 \rule[-2ex]{0pt}{3ex} Mumbai & 11:17 & 13:19 & 15:05 & 0.545 & 46 & 23& 36-63.8\checkmark\\ 
 \rule[-2ex]{0pt}{3ex} Shillong & 12:19 & 14:04 & 15:32 & 0.752 & 45 & 47& 10-20*, 12.9-63.6\checkmark, 56-62\dag  \\ 
 \rule[-2ex]{0pt}{3ex} Jaipur & 11:45 & 13:34 & 15:09 & 0.426 & 52 & 45& 10*, 12.8-60.1\checkmark \\ 
 \hline
 \multicolumn{8}{l}{\textmusicalnote\cite{Galav2010}*\cite{Vyas2012}, \dag\cite{Sanjaykumar2012}} \\
 \multicolumn{8}{l}{\ddag \cite{MadhavHaridas2012}, \checkmark\cite{Momani2010}}\\
 \multicolumn{6}{l}{Time is in IST(UT+5:30)}\\
 \multicolumn{6}{l}{CD1: Normal EEJ Control Day, CD2: Preceding Control Day}
  \end{tabular} 
  \end{table}
  \end{landscape}

 \begin{figure}[htb]
 \centering
 \includegraphics[angle=0, width=6.5in]{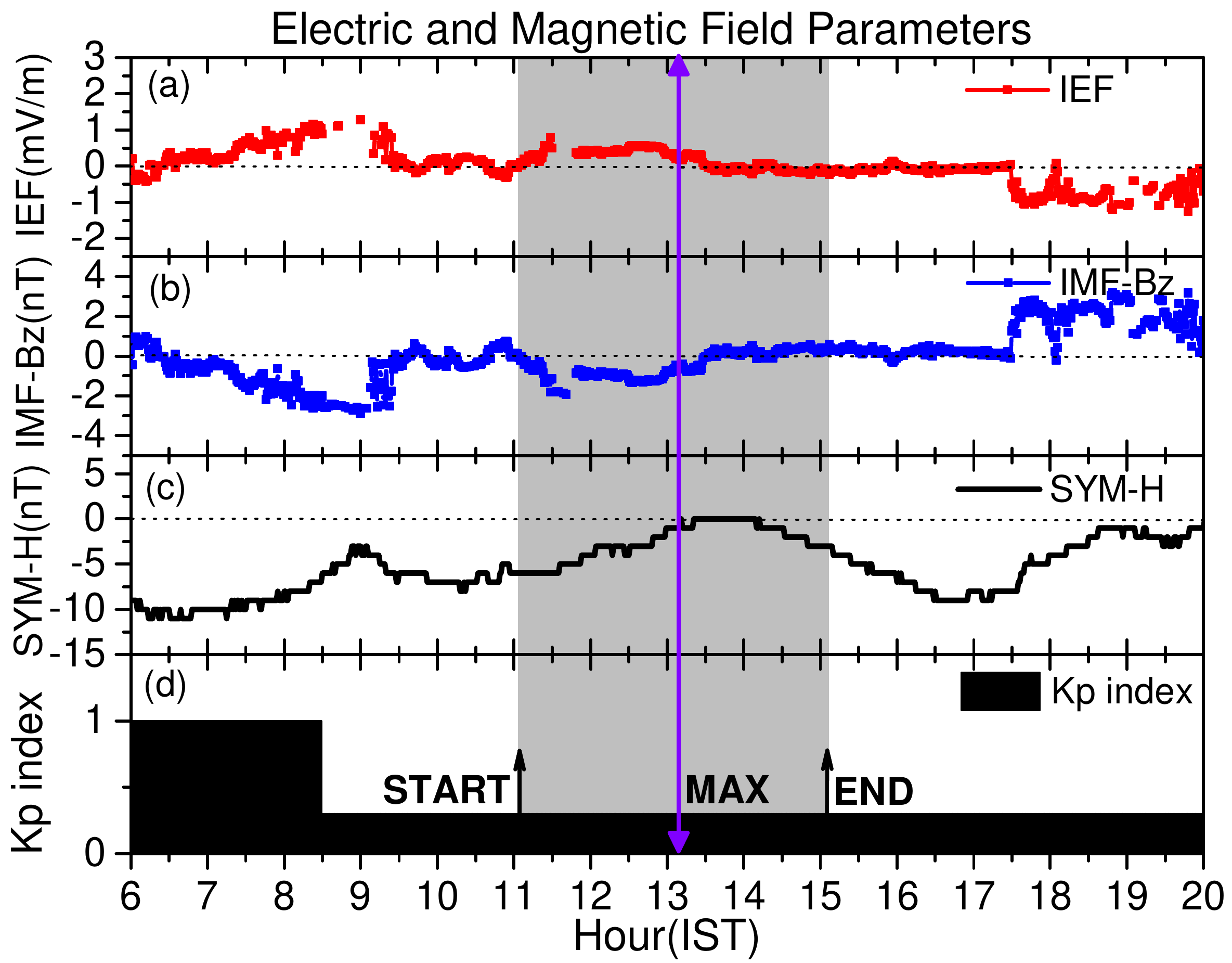}
 \caption{Interplanetary and geomagnetic parameters during 15 Jan 2010 annular solar eclipse. Panels a,b, c, and d depict interplanetary electric  field (IEF), interplanetary magnetic field (IMF-Bz), SYM-H index, and Kp index respectively. The shaded region outlines the eclipse duration, with arrows indicating start, maximum, and end of the eclipse.}
 \label{fig:geoindices1}
 \end{figure}
 
  \begin{figure}[htb]
   \centering
   \includegraphics[width=6.5in]{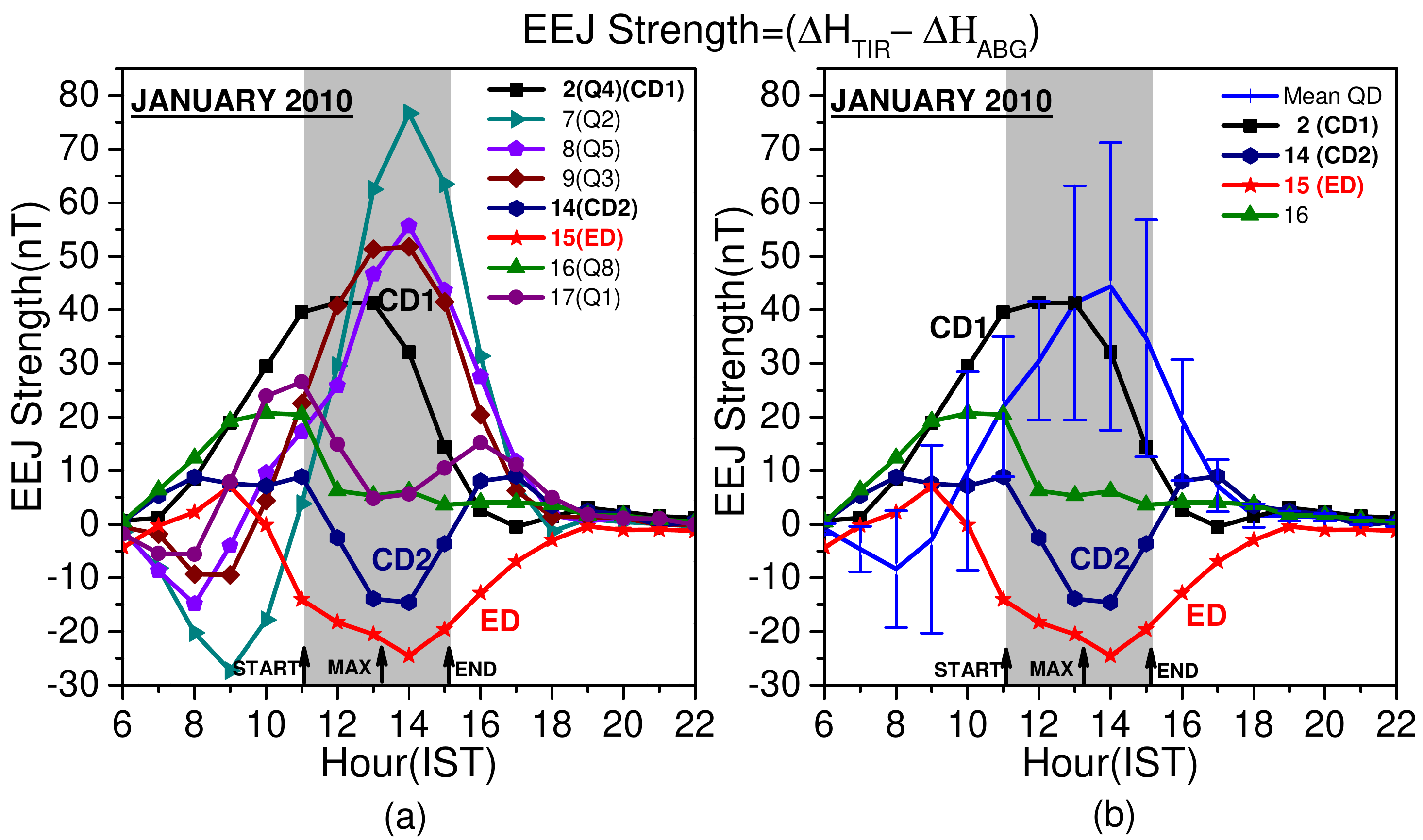}
   \caption{Equatorial electrojet (EEJ) strength over Indian region; (a) on the eclipse day (ED; 15 Jan 2010), possible control days, and 5 Q-Days of the month, (b) on the eclipse day (ED; 15 Jan 2010), its succeeding day (16 Jan 2010), control days (CD1: 2 Jan 2010, CD2: 14 Jan 2010), and average of 5 Q-Days of the month.}
   \label{fig:eejstrengths}
   \end{figure}

 \begin{landscape}
  \begin{figure}[htb]
  \begin{minipage}[a]{0.5\linewidth}
    \centering
  \includegraphics[width=.99\linewidth]{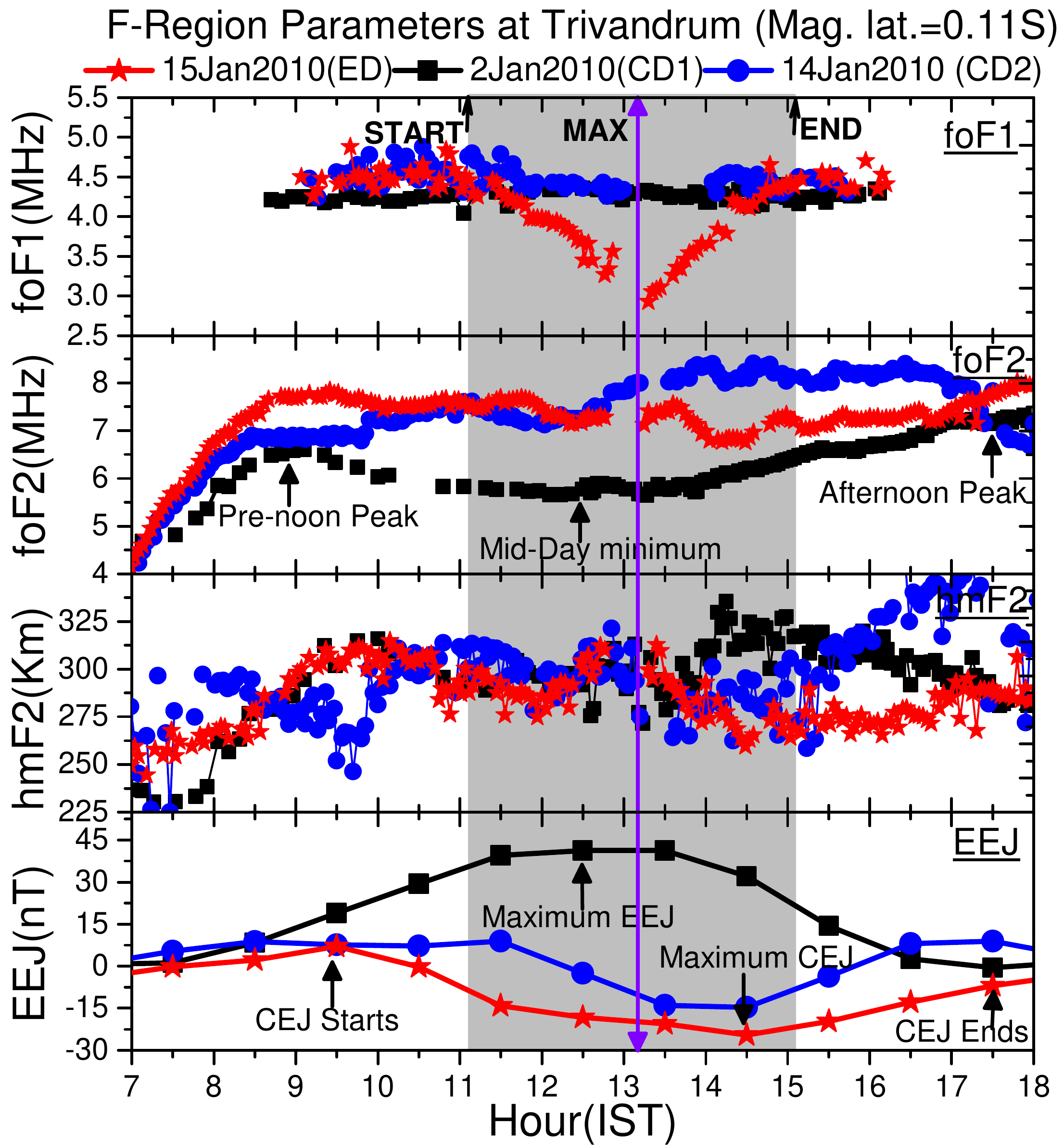}
    \centerline{(a)}\medskip
  \end{minipage}
  \hfill
  \begin{minipage}[a]{0.5\linewidth}
    \centering
   \includegraphics[width=.99\linewidth]{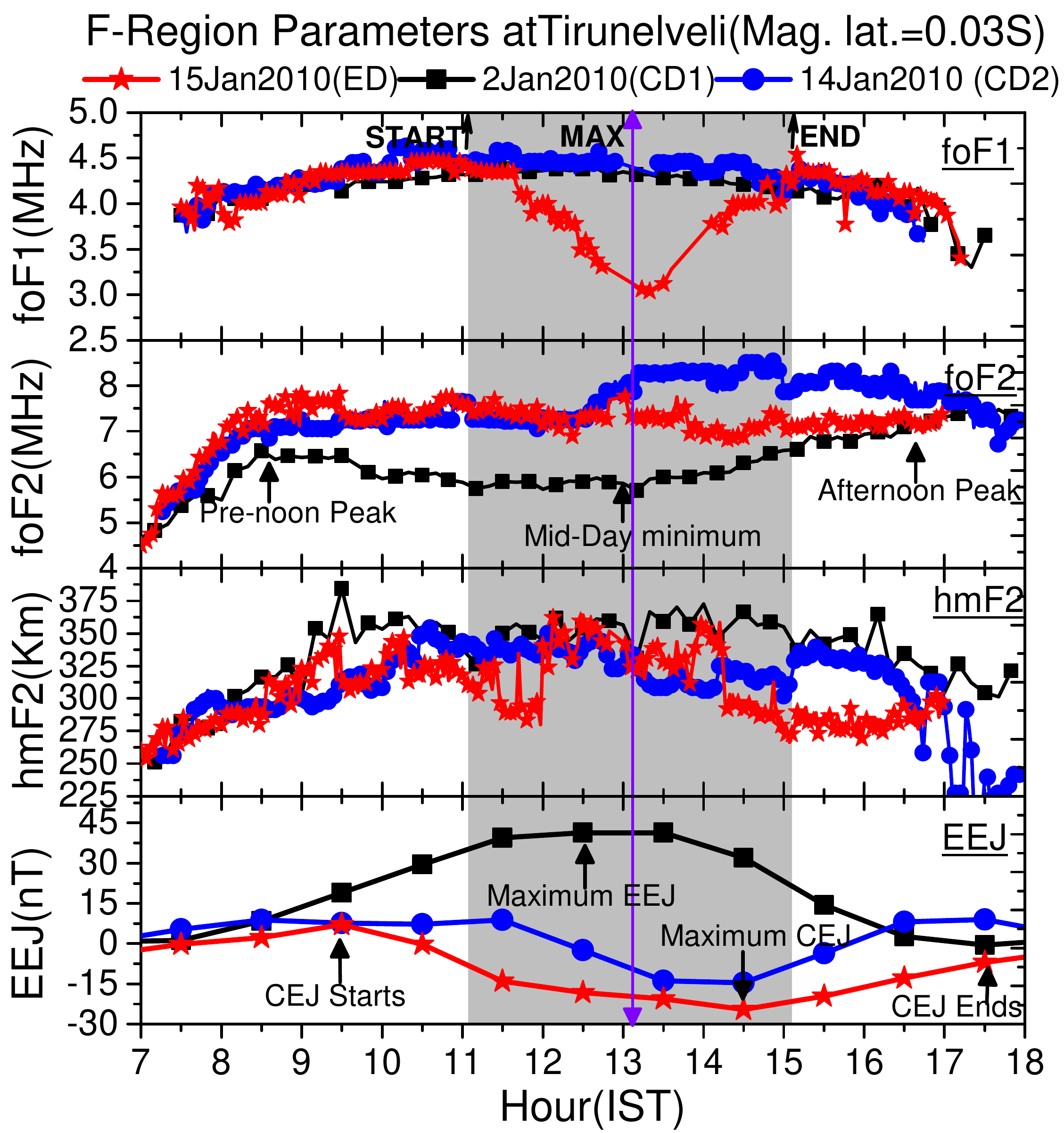}
    \centerline{(b)}\medskip
  \end{minipage}
  \caption{Temporal variations of foF1 (top panel), foF2 (second from top), hmF2 (second from bottom) and their comparison with the control days recorded from the ionosondes at (a) Trivandrum and (b) Tirunelveli. The lower panel in both the columns show the temporal variations of EEJ strength during the respective days. The shaded region outlines the eclipse duration, with arrows indicating start, maximum, and end of the eclipse. }
  \label{fig:ionosonde parameters}  %
  \end{figure}
  \end{landscape}

  \begin{figure}[htb]
  \begin{center}
  \includegraphics[angle=0,width=6.5in]{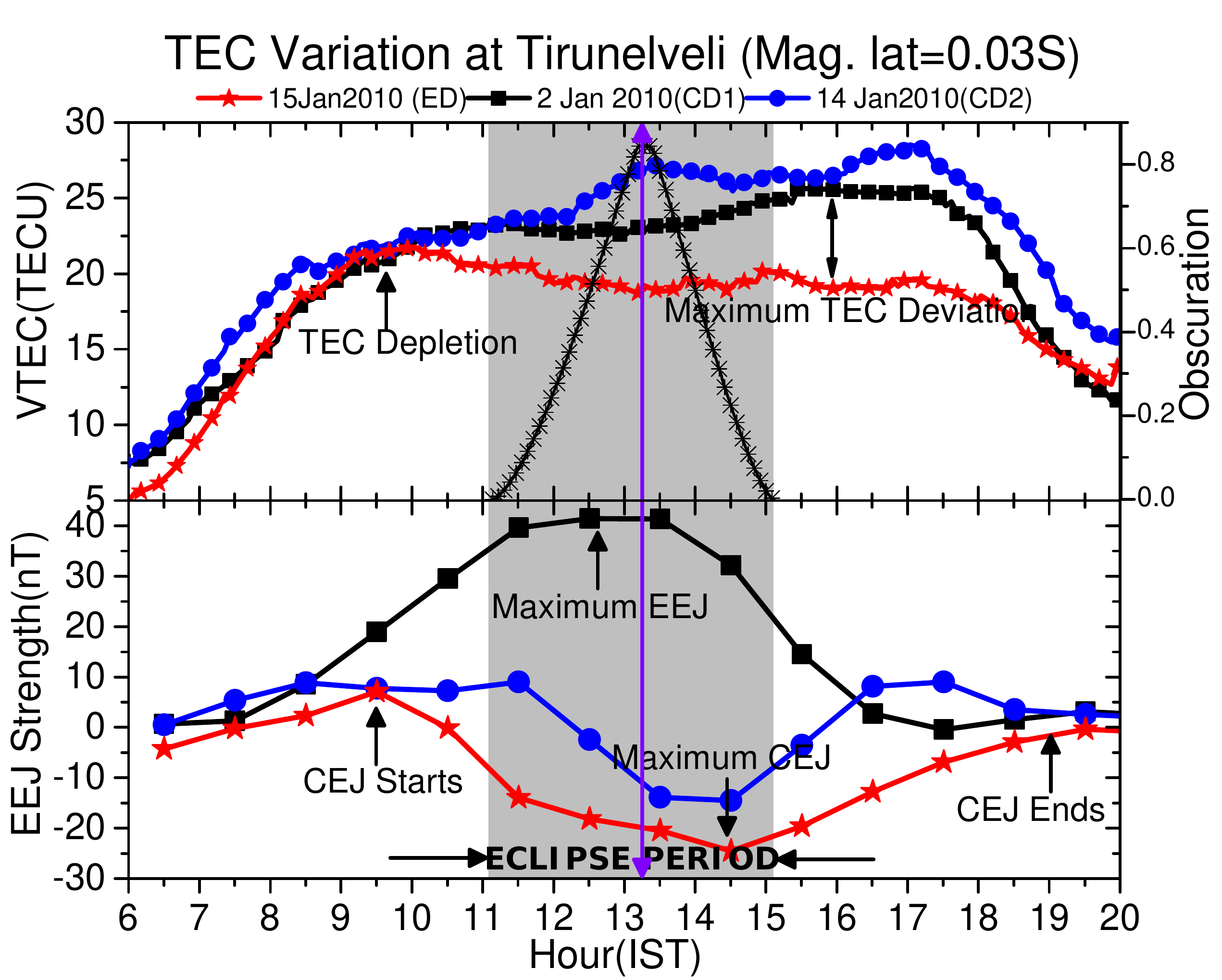}
  \caption{Temporal variations of VTEC and EEJ strength at Tirunelveli during the eclipse day (ED; 15 Jan 2010) and their comparisons with the normal EEJ control day (CD1; 2 Jan 2010) and the preceding control day (CD2; 14 Jan 2010). The black line with stars show the variations in obscuration magnitude. The shaded region outlines the eclipse duration with a two-headed arrow in the middle which signifies the centrality of the eclipse. The small black arrows indicate maximum amplitudes of EEJ/CEJ.}
  \label{fig:TECvariation_Tirunelveli}
  \end{center}
  \end{figure}
  
  \begin{figure}[htb]
  \begin{center}
  \includegraphics[angle=0,width=6.5in]{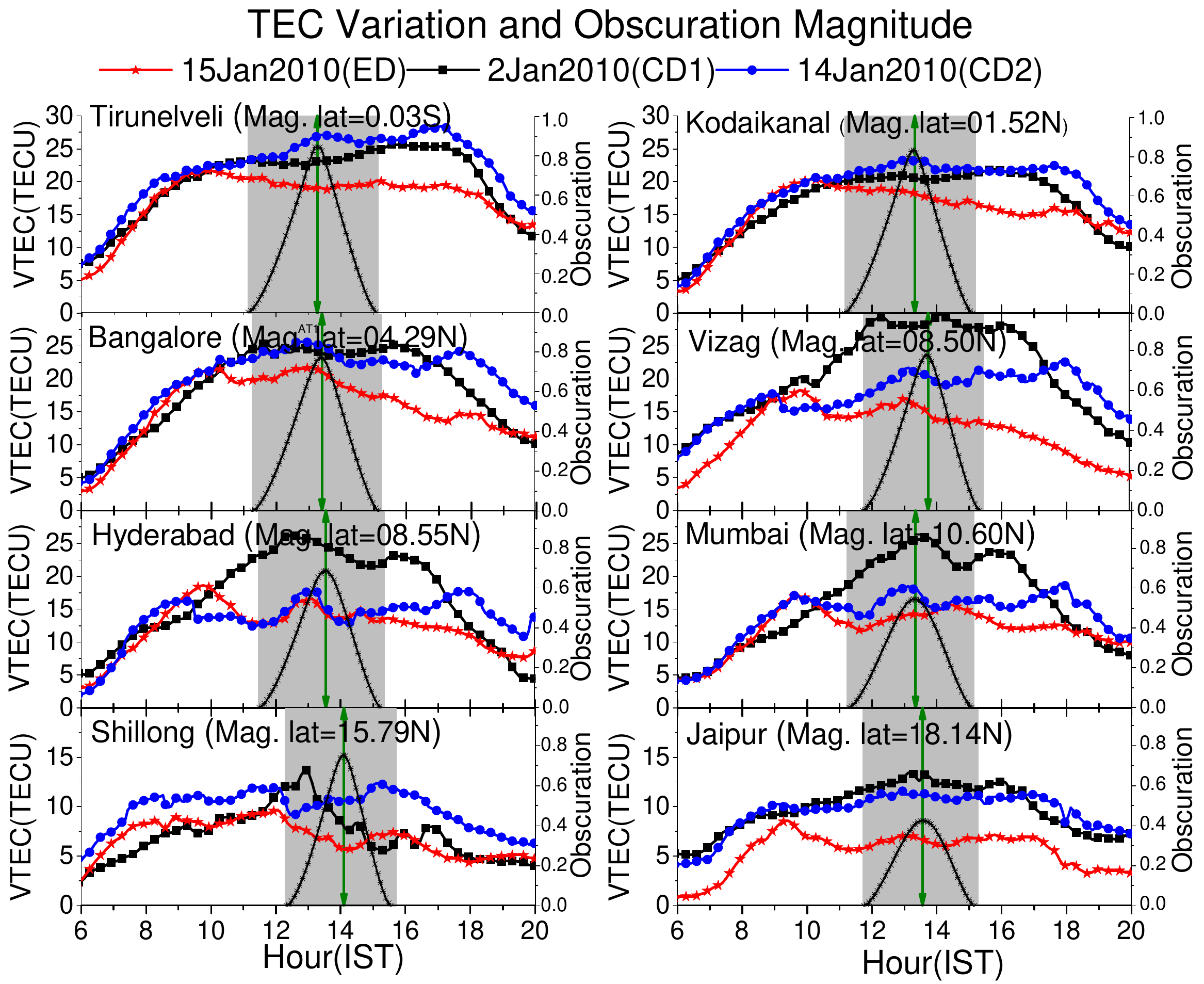}
  \caption{Temporal variations of VTEC and obscuration magnitude at different locations during the eclipse day (ED; 15 Jan 2010) and their comparisons with the normal EEJ control day (CD1; 2 Jan 2010), and the preceding control day (CD2; 14 Jan 2010). The black lines with stars show the variations in obscuration magnitude. The shaded regions outline the eclipse duration with the two-headed arrows in the middle which signify the centrality of the eclipse at each station.}
  \label{fig:AllstationTEC}
  \end{center}
  \end{figure}

  \begin{figure}[htb]
  \centering
  \includegraphics[height=6.8in,width=5in]{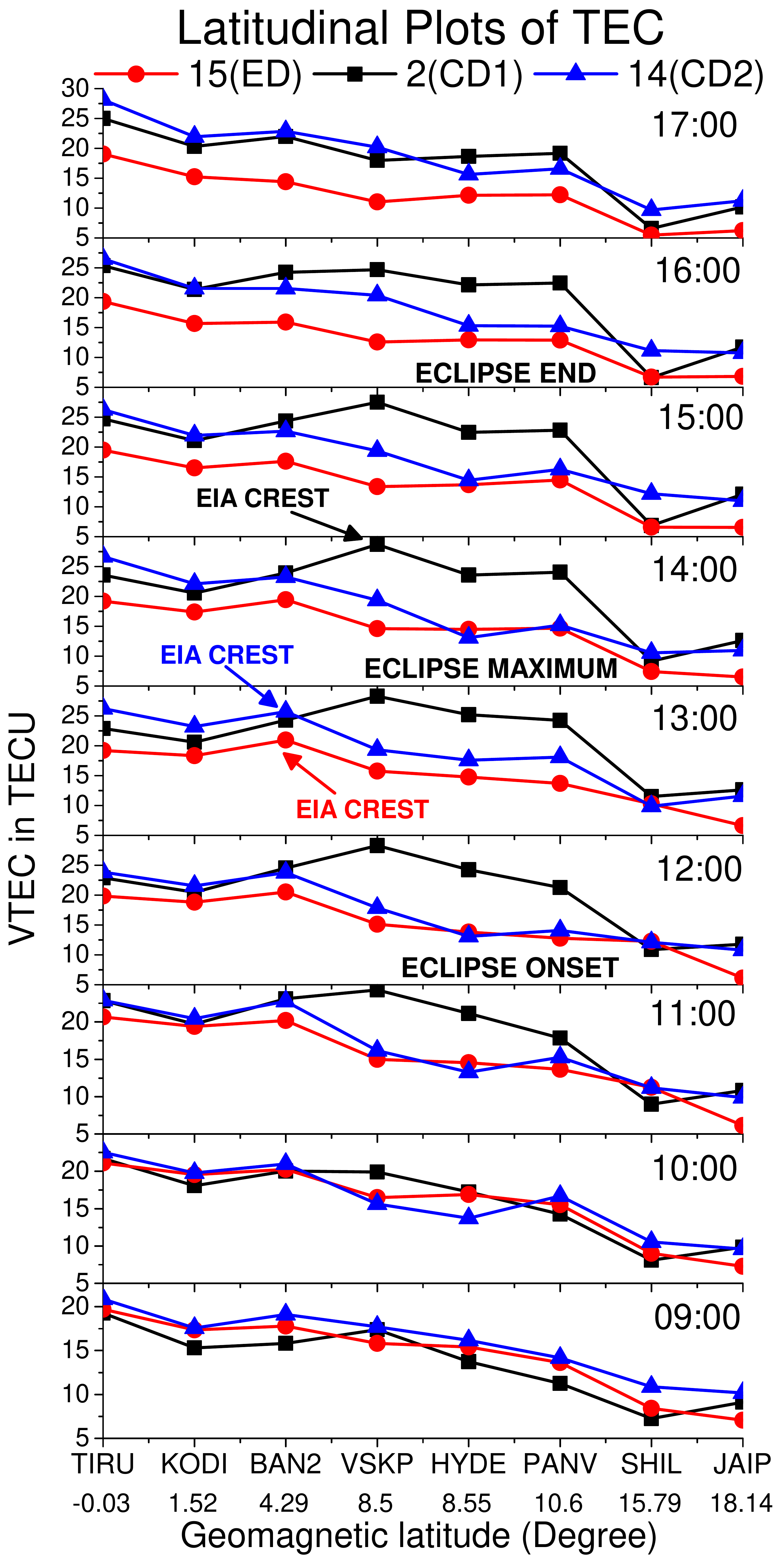}
  \caption{Latitudinal plots of hourly VTEC on the eclipse day (ED; 15 Jan 2010) and their comparisons with the normal EEJ control day (CD1; 2 Jan 2010), and the preceding control day (CD2; 14 Jan 2010). The arrows indicate latitudes of EIA crests on the respective days.}
  \label{fig:LatitudinalTEC}
  \end{figure}

 \begin{figure}[htb]
 \begin{center}
 \includegraphics[width=6.8in]{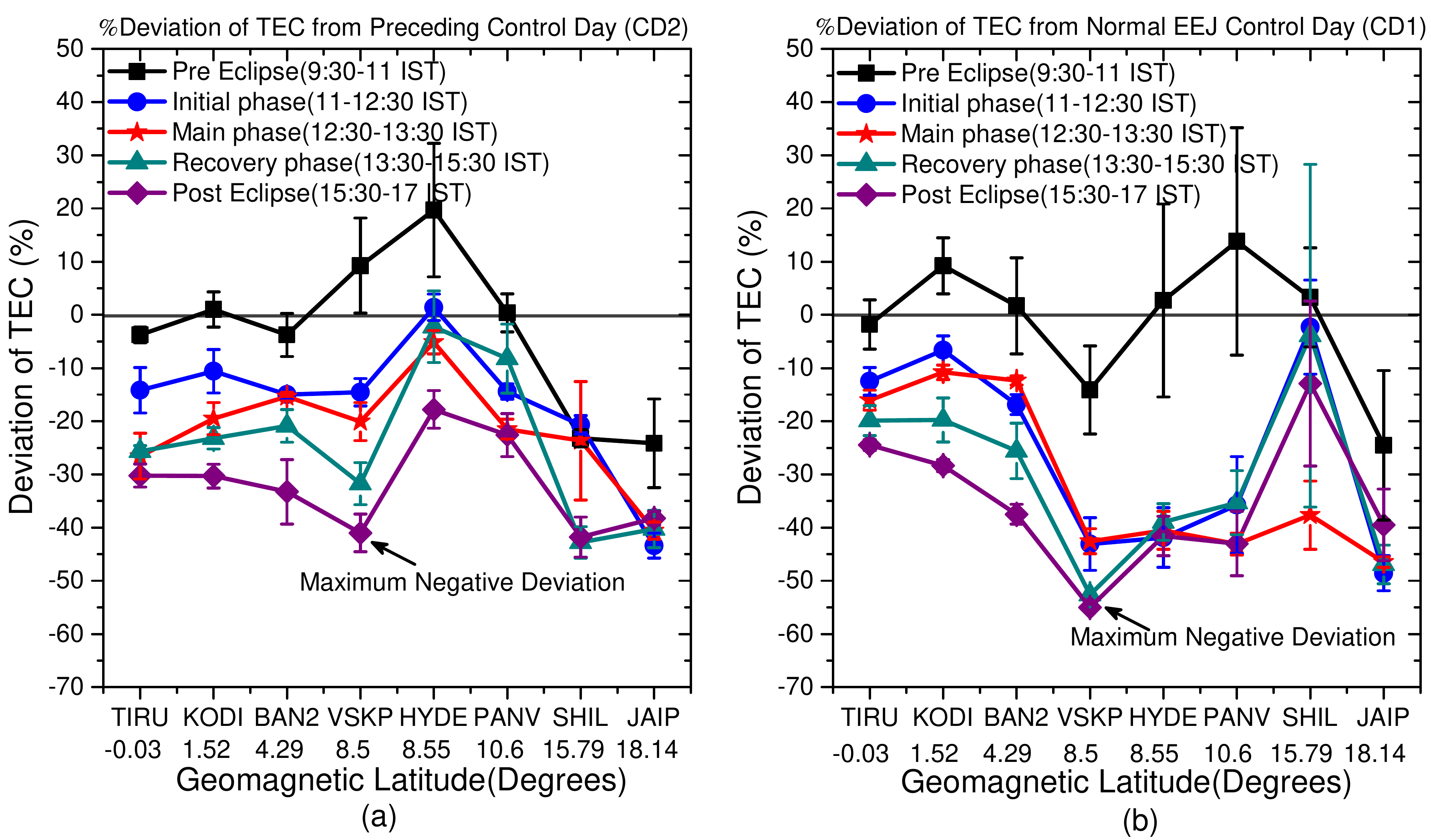}
  \caption{Latitudinal plots of percentage deviation of VTEC on the eclipse day (ED; 15 Jan 2010) and their comparisons with (a) the preceding control day (CD2; 14 Jan 2010) and (b) the normal EEJ control day (CD1; 2 Jan 2010) during the pre-, post- and the intermediate (initial, main and recovery) phases of the eclipse.}
 \label{fig:PercentagedeviationofTEC}
 \end{center}
 \end{figure}
 









\end{document}